\documentclass[12pt]{amsart}

\pdfoutput=1

\usepackage[text={500pt,650pt},centering]{geometry}
\usepackage{bbm}
\usepackage{enumitem}
\usepackage{url}
\usepackage{amssymb,amsfonts,amsmath,amsthm}
\usepackage{esint}
\usepackage{mathtools} 
\usepackage{graphicx}
\usepackage{MnSymbol}
\usepackage{mathtools} 
\usepackage[bottom]{footmisc}
\usepackage[colorlinks=true, pdfstartview=FitV, linkcolor=blue, citecolor=magenta, urlcolor=blue,pagebackref=false]{hyperref}
\usepackage{microtype}
\usepackage{cite}
\usepackage{xcolor}
\usepackage{bigints}

\usepackage{bm}
\usepackage{dsfont}   

\newenvironment{talign*}
 {\csname align*\endcsname}
 {\endalign}

\def \({\left(}
\def \){\right)}
\def \[{\left[}
\def \]{\right]}

\def\bx{\mathbf{x}}
\def\by{\mathbf{y}}

\newcommand{\bz}{{\mathbf{z}}}

\newcommand{\be}{\begin{equation}}
\newcommand{\ee}{\end{equation}}

\newcommand{\bea}{\begin{align}}
\newcommand{\eea}{\end{align}}

\DeclareMathAlphabet{\varmathbb}{U}{bbold}{m}{n}

\newcommand{\EE}{\mathbb{E}}

\begin{document}

\title[High-dimensional inference: a statistical mechanics perspective]{High-dimensional inference: \\a statistical mechanics perspective}

\author[J. Barbier]{Jean Barbier}
\address[Jean Barbier]{International Center for Theoretical Physics, Trieste, Italy.}
\email{jbarbier@ictp.it}

\date{}
\maketitle

\begin{abstract}
Statistical inference is the science of drawing conclusions about some system from data. In modern signal processing and machine learning, inference is done in very high dimension: very many unknown characteristics about the system have to be deduced from a lot of high-dimensional noisy data. This ``high-dimensional regime'' is reminiscent of statistical mechanics, which aims at describing the macroscopic behavior of a complex system based on the knowledge of its microscopic interactions. It is by now clear that there are many connections between inference and statistical physics. This article aims at emphasizing some of the deep links connecting these apparently separated disciplines through the description of paradigmatic models of high-dimensional inference in the language of statistical mechanics.

This article has been published in the issue on artificial intelligence of Ithaca, an Italian popularization-of-science journal. The selected topics and references are highly biased and not intended to be exhaustive in any ways. Its purpose is to serve as introduction to statistical mechanics of inference through a very specific angle that corresponds to my own tastes and limited knowledge.
\end{abstract}

\section{Statistical inference: old and new}

\textbf{Statistical inference} aims at accurately describing some system, through the design of an appropriate probability distribution, based on data about this system and, potentially, some assumptions about it. Designing both statistically and computationally efficient inference procedures is thus crucial in virtually all fields of science.

Classical statistics is mostly concerned with the regime where the system of study is rather ``simple'', or ``low-dimensional''. Namely, it is parametrised by few quantities of interest and the amount of accessible data is large. But in the \textbf{big-data era}, contemporary signal processing and machine learning tasks require performing inference in the so-called \textbf{high-dimensional (high-d) regime}. This means that even if the amount of data as well at its dimensionality may be (very) large, the number of unknown parameters characterizing the system under study is also huge. Therefore totally new statistical tools are required to make sense of the data in order to ``extract the signal from the noise''.

\subsection{Classical ``low dimensional'' statistics}

In classical statistics the \textbf{signal}, namely the information of interest/unknown parameter to recover from the data, is \textbf{low-dimensional}. To be more precise, let us denote the signal $\bx\in \mathbb{R}^p$ and the data $\by=\by(\bx)\in \mathbb{R}^n$, that depends on the signal. For example consider a simple experiment where one tries to infer if a coin is fair, namely, whether $\mathbb{P}(\text{head})=x=1-\mathbb{P}(\text{tail})$ with $x=1/2$. In this example the parameter space has dimension $p=1$ as the relevant parameter/signal $x\in[0,1]$ is a scalar. A natural protocol to answer that question is: toss the coin $n$ times and record the number $n_{\text{h}}\in\{0,\ldots,n\}$ of times it fell on head. Then in the limit $n\gg 1$ the \textbf{law or large numbers}, which is a fundamental statistical property at the core of the apparent predictability of the world in spite of its inherent probabilistic nature, predicts that the empirical mean converges to the statistical mean. This translates here to $\frac{1}{n}\sum_{i=1}^n \mathbbm{1}(\text{toss}_i=\text{head})=n_{\text{h}}/n = x +o_{n}(1)$ with a correction $o_n(1)\to 0$ as $n\to \infty$ (here $\mathbbm{1}(\cdot)$ is the indicator function). Less basic examples could be to infer the earth gravitational constant from the recording of $n\gg 1$ trajectories of falling objects with various initial conditons (in which case again $p=1$), or infering the average height $x_1$, weight $x_2$ and the associated variances $(x_3,x_4)$ based on some large population of $n$ individuals; in the latter case $p=4$.

What is really important in these examples is that $$\frac pn \ll 1$$ is very small. In this regime the optimal way to infer the parameters from the data, using a probabilistic model for how the data is generated conditional on the unknown paramaters $\bx$, is through \textbf{maximum likelihood estimation (MLE)}. The probabilistic way to represent a random process of data generation is a probability distribution of observing the data conditional on the (unknown) parameters $\mathbb{P}(\by\mid \bx)$, called \textbf{likelihood}. For example, in the 
coin tossing experiment, an obvious observation is that conditional on the bias parameter $x$ all tosses are independent. Therefore, mapping the binary data-space $\{\text{head},\text{tail}\}$ to $\{0,1\}$, each toss is a Bernoulli experiment with $\mathbb{P}(y_{i}=0\mid x)=x$. Therefore the likelihood that the random variable (r.v.) $N_{\text{h}}$ takes value $k\in\{0,\ldots,n\}$, $N_{\text{h}}$ being the random number of heads among $n$ trials, is the binomial law of (unknown) success probability $x$: $\mathbb{P}(N_{\text{h}}=k\mid x)={n \choose k}x^k (1-x)^{n-k}$. In this experiment the only data is the outcome $n_{\text{h}}$ of $N_{\text{h}}$ as the order of the tosses is irrelevant. Therefore $\mathbb{P}(N_{\text{h}}=n_{\text{h}}\mid x)$ is the likelihood of the data.

It is conceptually useful to introduce the \textbf{likelihood function} $$\mathcal{L}(\bx\mid \by):=\mathbb{P}(\by\mid \bx).$$ It is important to think of $\mathcal{L}(\bx\mid \by)$ really as a function of the parameters given the data (the data being fixed and cannot be modified), not the other way around as suggested by the conditional probability distribution $\mathbb{P}(\by\mid \bx)$. This is the reason behind the introduction of a specific notation $\mathcal{L}(\bx\mid \by)$ emphasising this correct interpretation. MLE says that one should take as estimate $\hat{\bx}=\hat{\bx}(\by)$ of the unknown parameters $\bx$ the value that maximizes the (logarithm of the) likelihood function given the measured data. In coin tossing $\mathcal{L}(x\mid n_{\text{h}}):=\mathbb{P}(N_{\text{h}}=n_{\text{h}}\mid x)$, so MLE gives
\begin{align}
\hat x&\in \underset{{x\in[0,1]}}{\text{argmax}} \ln\mathcal{L}(x\mid n_{\text{h}}) = \underset{{x\in[0,1]}}{\text{argmax}}\{n_{\text{h}}\ln x + (n-n_{\text{h}})\ln(1-x)\}.  \nonumber
\end{align}
The function $\{\cdots\}$ is concave so its unique maximiser is easily found to be $\hat x(n_{\text{h}})=n_{\text{h}}/n$. The principled approach of MLE therefore allows to recover the natural choice suggested by the law of large numbers. This simple but powerful method has driven classical statistics for more than a century since its development by C.~Gauss, P.~S.~Laplace or F.~Edgeworth. Finally, MLE can be shown to be optimal in the classical limit of statistics $p/n\to 0$ in a precise and quite general sense\footnote{The MLE estimator $\hat{\bx}$ is optimal in the following sense: in the Bayesian optimal setting described soon, in the regime $n\gg p$, $\hat{\bx}$ is both a Bayes estimator (namely, it minimizes the Bayes risk associated with the distribution of the signal $\mathbb{P}(\bx)$) and minimax for the mean-square loss, see the upcoming section on decision theory and Chapter 12 in the great book \cite{wasserman}.}.

\subsection{High-dimensional statistics}
The high-d regime generally refers to statistical settings in which both the number of parameters and of data points --that can themselves be high-d-- are large and comparable: $p/n \to \delta >0$ as both $p,n\to \infty$ together, with $\delta$ an order one constant. This is to be contrasted with the classical limit $\delta \to 0$. To be more precise we require 
\begin{align}\label{highdregime}
\frac{p}{n \times \text{SNR}_{\text{d}}} \to \delta >0,	
\end{align}
where the \textbf{signal-to-noise ratio (SNR)} per data point $\text{SNR}_{\text{d}}$ is a measure of the information content about $\bx$ carried by a single data point, in average. The definition of $\text{SNR}_{\text{d}}$ depends on the model under study, but is always related to a natural way of comparing the (average) signal amplitude with the one of the noise that corrupts the data. When it is of order one we recover the usual definition of high-d regime $p/n \to \delta >0$.

In the coin tossing experiment, a natural way of quantifying the noise corrupting the single data point $y=n_{\text{h}}\in \{0,\ldots,n\}$ is given by the variance of the associated r.v. $N_{\text{h}}$. The variance of the binomial distribution $\text{Bin}(n,x)$ is $nx(1-x)$ so $\text{SNR}_{\text{d}}=nx(1-x)$ is large. Therefore  $\# \text{paramaters}\div (\# \text{data points} \times \text{SNR}_{\text{d}}) = 1/(1\times nx(1-x)) \to 0$ as $n\to \infty$. An equally valid interpretation is: we have $n$ data points $(y_i)_{i=1}^n$, each one being the outcome of a Bernoulli experiment. Each $y_i\in\{0,1\}$ has variance $x(1-x)=O(1)$. Then $\# \text{paramaters}\div (\# \text{data points} \times \text{SNR}_{\text{d}}) = 1/(n\times x(1-x)) \to 0$. Because it tends to $0$ we are in the classical regime of statistics.

The modern statistical regime \eqref{highdregime} is particularly relevant for applications in all sorts of signal processing tasks --image and sound processing, medical applications, error-correcting codes for communications, etc-- and machine learning --automatic image classification, drug discoveries, natural language processsing and translation, self-driving cars, etc-- that are changing the world at a unprecedented paste. One example are the modern \textbf{deep neural networks} trained on data-bases with millions images. But the number of parameters/synaptic weights defining these complex models is of the same order, or even much bigger. See, e.g., \cite{highbias} for an introduction to machine learning for physicists.

It is not exagarating to call that a \textbf{data-revolution}, and high-d statistics is the theoretical powerhouse at its core. The other key pillars of this revolution being the amount of accessible data, as well as the modern computers and specific computational units able to process such huge data sets, like graphical processing units (GPUs). 

Understanding the high-d regime requires totally new concepts and mathematical tools, and for solving actual applied problems, we need new algorithms. Speaking about algorithms, in the high-d world where the data is so massive, not only \textbf{statistical efficiency} matters --namely, the capacity of an algorithm to extract the relevant information, independently of any ``speed concern''--, but also \textbf{computational efficiency}, as it quickly becomes a bottleneck. Researchers working in applications of high-d statistics must always keep in mind these two considerations, a key feature of the field that makes it so interesting and challenging at the same time. 

In this article we will mainly focus our attention on the \textbf{information-theoretic} (or \textbf{statistical}) limitations to inference and leave the algorithmic considerations aside.

\section{Basics of Bayesian inference}

\subsection{Breaking the curse of dimensionality using a-priori knowledge}

We have said that in classical statistics MLE estimation is optimal. This is not true anymore in the high-d regime. Because the amount of data is comparable to the number of unknown parameters to infer, this may create degeneracies in the solution of MLE, in the sense that the set $\text{argmax}_{\bx\in{\mathbb{R}^p}}\ln\mathcal{L}(\bx\mid \by)$ may have a huge cardinal, all solutions being equally bad. This is one issue related to the \textbf{curse of dimensionality}: in the high-d regime the volume of the space in which lives the signal $\bx$ increases so fast (exponentially fast with $p$) that the available data becomes relatively sparse. This sparsity is problematic for any method that requires statistical significance. In order to obtain a statistically sound and reliable result, the amount of data needed to support the result often grows exponentially with the dimensionality, but such an amount is never accessible (compute $\exp p$ with $p=10,100,1000$...). 

Therefore one has to fill-in this ``information gap'' due to relative lack of data using \textbf{assumptions} about $\bx$, namely, \textbf{a-priori knowledge}. Quoting D.~Mackay: ``you cannot do inference without making assumptions'', see the amazing book \cite{mackay}. This can be formalised through a probability distribution $\mathbb{P}(\bx)$ that depends on the signal only, and therefore is completely independent of the data $\by$. This distribution, called \textbf{prior}, translates in the language of probability the whole set of a-priori assumptions made by the statistician about $\bx$, before that the data is collected. It is crucial that once the data is acquired, the prior is not modified accordingly, otherwise this may create \textbf{bias}. Such assumptions could be, for example, that the signal is binary $\bx\in\{-1,1\}^p$ with uniform independent and identically distributed (i.i.d.) components $x_i$ (like the bits received from some communication source). This very basic assumption would translate in $\mathbb{P}(\bx)=\prod_{i=1}^p \frac12(\delta_{x_i,-1}+\delta_{x_i,1})$. But maybe in addition one may know that actually the signal is \textbf{sparse}\footnote{The sparsity assumption allows to reconstruct very high-d signals from relatively few data points, and, e.g., to invert apparently under-determined systems of linear equations. This forms the basis of a whole field of research in mathematics and signal processing called \textbf{compressive sensing}, see the excellent introduction \cite{CS} or \cite{FloCS} for a statistical mechanics approach.}, meaning it has a fraction $\rho$ of entries equal to $0$. In this case $\mathbb{P}(\bx)=\prod_{i=1}^p \{\rho \delta_{x_i,0}+\frac{1-\rho}{2}(\delta_{x_i,-1}+\delta_{x_i,1})\}$. And so on: the richer the set of assumptions, the more complex is the prior.  

\subsection{Combining assumptions and data: Bayes formula}

One of the most elegant probabilistic statement ever is the so-called \textbf{Bayes formula}. It has been discovered by a reverend named T.~Bayes before his death in 1761. Independently of Bayes, P.~S.~Laplace formalised similar ideas in 1774. Jeffreys, one of the father of modern Bayesian statistics, later wrote that Bayes's theorem "is to the theory of probability what the Pythagorean theorem is to geometry". As simple as it is, it encapsulates in a single equation a powerful recipe, more useful than ever in the context of inference in high dimensions. It says that the proper way to combine a-priori knowledge and data is through a simple multiplication followed by a normalization:
\begin{align}\label{Bayes}
\mathbb{P}(\bx\mid \by) =\frac{\mathbb{P}(\bx)\mathbb{P}(\by\mid \bx)}{\mathbb{P}(\by)}.
\end{align}
That is, $\text{posterior} = \text{prior}\times \text{likelihood}/\text{evidence}$. The \textbf{posterior distribution} $\mathbb{P}(\bx\mid \by)$ signifies 
\begin{align*}
\mathbb{P}(\text{the parameters take value}\ \bx \ \text{given the data is} \ \by \ \text{and our a-priori knowledge}).	
\end{align*}
``Posterior'' is in the sense of a-posteriori that the data has been collected. The posterior distribution is therefore the multiplication of the prior, that formalises all our assumptions about the signal, with the likelihood, that models the data-generating process conditional on the signal. The posterior combines in a single probability distribution all information we have about the signal, as well as our uncertainty about it through, e.g., the posterior variance $$\text{Var}(\bx \mid \by)=\mathbb{E}\big[\|\bx - \mathbb{E}[\bx\mid \by] \|^2\mid \by\big]$$ (defining generically $\mathbb{E}[g(\bx)\mid \by]:=\int d\bx \,g(\bx) \,\mathbb{P}({\bx}\mid {\by})$).

The normalization $$\mathbb{P}(\by)=\int d\bx'\,\mathbb{P}(\bx')\mathbb{P}(\by\mid \bx')$$ is called \textbf{evidence}. It is the marginal distribution of the data. Note that in high dimensions this distribution can be very hard, if not impossible, to compute exactly as it requires to perform a $p$-dimensional integral, with $p$ very large. This is often the main computational bottleneck in Bayesian inference and there is a whole field of research related to finding good approximation techniques for the evidence.

\subsection{Information-theoretic limits: the Bayesian optimal setting}

From now on we will restrict our discussion to the \textbf{Bayesian optimal setting}. This means that the statistician knows the model underlying the data-generating process (but of course not the signal $\bx$), which translates into the correct likelihood $\mathbb{P}(\by\mid \bx)$. In addition she also exploits correctly all a-priori information about the signal, namely, the signal has indeed been randomly generated from the prior $\mathbb{P}(\bx)$ used by her. Therefore in this setting, the posterior distribution is the ``correct one'' and all the upcoming discussion applies only to this case. 

The Bayesian optimal setting is fundamental: as we will argue, any \textbf{optimal estimator}, whatever meaningful notion of optimality is considered, relies on the correct posterior. Therefore in order to study the \textbf{information-theoretical limits} of inference --namely the best results one can aim for independently of any computational concern--, one needs to precisely study inference in this optimal setting. The performance of estimators for $\bx$ in this setting cannot be outperformed by any algorithm, even those allowed to run for infinite time. The Bayesian optimal setting is at the core of \textbf{information theory} \cite{cover1999elements}.

The mismatched setting where the likelihood used is not properly describing the data generation and/or the prior is biased (i.e., does not correspond to the probability distribution from which the signal was generated) is much more complicated and goes beyond the present discussion. In the repeated coin tossing experiment, a wrong a-priori assumption could be that the tosses are independent, while for some reason the tosses were correlated; this could be due, e.g., to a tiny demon living inside the coin and that would modify the probability $x=x_i(y_{i-1})$ of head for the toss $i$ as a function of the previous tossing result $y_i$. Yet, a lot of the phenomenology, that is inherent to the high-dimensionality, remains the same in the Bayesian optimal and mismatched (more realistic) settings.

\subsection{Optimality of estimators: Bayesian decision theory}

In order to quantify the performance of the statistician in estimating $\bx$, we need to define a proper recontruction error associated with a given estimator $\hat{\bx}=\hat{\bx}({\by})$. This error metric is called \textbf{loss} in statistics, and can be thought as an energy function. There are many possible choices, whose relevance depends on the specific application at hand. One canonical choice is the $0-1$ loss: $\ell(\hat{\bx},\bx)=1-\mathbbm{1}(\hat{\bx}= \bx)$. This choice makes sense when the signal is discrete, like in communications where the signal is made of bits. The loss cannot be computed as it depends on the unknown signal. Therefore in order to define a notion of ``goodness'' of an estimator we define the \textbf{posterior risk} (our best estimate of the loss):
\begin{align*}
r(\hat{\bx}\mid {\by}):=\int d{\bx} \, \mathbb{P}({\bx}\mid {\by})\ell(\hat{\bx},\bx),
\end{align*}
which is simply equal to $1-\mathbb{P}(\hat{\bx}\mid \by)$ for the $0-1$ loss, or its average with respect to the evidence, called \textbf{Bayes risk} $$r(\hat{\bx}):=\int d\by\,\mathbb{P}(\by)r(\hat{\bx}\mid {\by}).$$ Both can be in theory computed from the knowledge of the data only and the underlying statistical model; in practice this may be computationally very demanding in high dimension. We directly get that the optimal estimator, optimal in the sense of minimizing the posterior (or Bayes) risk, is in the case of $0-1$ loss given by the posterior mode: 
\begin{align*}
 \hat{\bx}_{\text{MAP}}(\by)&:=\underset{\,\hat{\bx}\in\mathbb{R}^p}{\text{argmin}}\, r(\hat{\bx}\mid {\by})=\underset{\hat{\bx}\in\mathbb{R}^p}{\text{argmax}}\, \mathbb{P}(\hat{\bx}\mid {\by}) .
\end{align*}
MAP stands for \textbf{maximum a-posteriori} estimator. Another common choice that is more appropriate for real-valued signals is the $L_2$ loss: $\ell(\hat{\bx},\bx)=\|\hat{\bx}- \bx\|^2$. The associated posterior risk is called the \textbf{mean-square error}, and the estimator that minimizes it is the \textbf{minimum mean-square error (MMSE) estimator}:
\begin{align*}
  \hat{\bx}_{\text{MMSE}}(\by)&:=\underset{\,\hat{\bx}\in\mathbb{R}^p}{\text{argmin}}\, \int d{\bx} \, \mathbb{P}({\bx}\mid {\by})\|\hat{\bx}- \bx\|_2^2 = \int d\bx \,\bx \,\mathbb{P}({\bx}\mid {\by})=:\mathbb{E}[\bx\mid \by].
\end{align*}
The second equality is easily shown by equating the gradient w.r.t. $\hat{\bx}$ of the posterior risk to the all-zeros vector (by convexity it leads the unique minimizer). Therefore the MMSE estimator is ``simply'' the posterior mean. Of course in general this may be very costly to compute because there are two $p$-dimensional integrals: the evidence (necessary to normalize the posterior), and then the integral $\int d\bx \,\bx \,\mathbb{P}({\bx}\mid {\by})$. Therefore in many practical applications one prefers the MAP estimator as it bypasses this issue (but is sub-optimal for any other loss than the $0-1$ loss). The principle that for a given (natural) loss/risk the associated optimal estimator relies on the posterior is general.

Focusing on the $L_2$ loss, the inference error associated with the MMSE estimator is, naturally, the \textbf{miminim mean-square error}:
\begin{align}
\text{MMSE}_p&:=\frac1p \|\mathbb{E}[\bx\mid \by]-\bx\|^2, \qquad \text{MMSE}:=\lim_{p}\mathbb{E}_{\bx,\by} \text{MMSE}_p = \lim_p \mathbb{E}_{\by} \text{Var}(\bx \mid \by).\label{MMSE}
\end{align}
By $\lim_{p}$ we always mean the ``thermodynamic limit'' $p\to\infty$. We consider the average error, the symbol $\mathbb{E}$ meaning the expectation with respect to the signal $\bx$ and the data $\by$ (conditional on $\bx$), seen as r.vs. drawn from their respective distributions (prior and likelihood). This quantity summarizes in a single number all the complexity of the high-d problem, by providing the optimal error one can aim for any algorithm, in average over all possible realisations of the problem.

One may wonder whether the number MMSE is sufficient to describe the problem, as it might a-priori be the case that the r.v. $\text{MMSE}_p$ (random through $(\bx,\by)$) fluctuates a lot (i.e., has $O(1)$ variance). But in well-defined high-d inference problems in the Bayesian optimal setting it \textbf{concentrates}; it is said to be \textbf{self-averaging} in physics terminology. This means that it actually does not fluctuate much for large systems, and becomes deterministic in the limit $p,n\to \infty$: 
\begin{align*}
\text{MMSE}_p=\text{MMSE}+o_p(1)
\end{align*}
where $o_p(1)$ is by definition a quantity such that $\lim_{p} o_p(1)=0$. Therefore it is sufficient to focus on the asymptotic averaged MMSE in order to capture/predict the behavior of fixed large (typical) instances of the problem. The self-averaging of error metrics in high-d Bayes-optimal inference is very generic \cite{barbierOverlap,barbierPanchenko} (in mismatched settings this is not necessary the case). It is a non-trivial manifestation of the phenomenon of \textbf{concentration of measure} in high-d models, which is at the core of the determinism/predictability of these complex random systems.

\section{High-dimensional inference as statistical mechanics}

Let us establish now clear connections between what we discussed until now on high-d inference and statistical mechanics.

\subsection{The posterior as a Gibbs-Boltzmann distribution}

The problem that we already mentionned of normalizing a high-d probability distribution like the posterior (i.e., computing the evidence) should ring a bell for physicists: the very same thing happens in statistical mechanics, where one of the main task is to compute the \textbf{partition function} $$\mathcal{Z}(\mathbf{J}):=\sum_{\boldsymbol{\sigma}}\, \exp\{-\beta \mathcal{H}(\boldsymbol{\sigma};\mathbf{J})\}$$ (if $\sigma_i$ is binary the sum is over $\boldsymbol{\sigma}\in\{-1,1\}^p$) which normalizes the \textbf{Gibbs-Boltzmann distribution}:
\begin{align}\label{GB}
\mathbb{P}_{\text{GB}}(\boldsymbol{\sigma};\mathbf{J})	=\frac{1}{\mathcal{Z}(\mathbf{J})}\exp\{-\beta \mathcal{H}(\boldsymbol{\sigma};\mathbf{J})\}.
\end{align}
Here $\mathcal{H}(\boldsymbol{\sigma};\mathbf{J})$ is the \textbf{Hamiltonian/energy function} defining the model, and $\boldsymbol{\sigma}$ (often binary) are the \textbf{spins}. $\mathbf{J}$ is the \textbf{quenched randomness}, namely, a set of fixed variables that parametrise the Hamiltonian. $\beta$ is the \textbf{inverse temperature}.

We can push the analogy further: the posterior distribution given by the Bayes formula \eqref{Bayes} can be naturally thought as a Gibbs-Boltzmann distribution in the context of high-d inference. Simply re-write it in exponential form and identify the (possibly real-valued) variables $\bx$ representing the unknown signal with the spins $\boldsymbol{\sigma}$, and the data $\by$ with the quenched randomness $\mathbf{J}$:
\begin{align*}
\mathbb{P}(\bx\mid \by) = \frac{1}{\mathcal{Z}(\by)}\exp\big\{\ln \mathbb{P}(\bx)+\ln \mathbb{P}(\by\mid \bx)\big\}.	
\end{align*}
So the partition function is the evidence, and the Hamiltonian is (minus) the log-prior plus log-likelihood, while $\beta=1$. In many interesting models the prior factorises over the signal entries $$\mathbb{P}(\bx)=\prod_{i=1}^p P(x_i)$$ (i.e., they are i.i.d.), and the likelihood factorises too over the data points $$\mathbb{P}(\by\mid \bx)=\prod_{j=1}^n Q(y_j\mid \bx)$$ (i.e., the data points are conditionally i.i.d.). In this case we recover a familiar form for the posterior:
\begin{align*}
\mathbb{P}(\bx\mid \by)=\frac{1}{\mathcal{Z}(\by)}\exp\big\{\sum_{i=1}^p\ln {P}(x_i)+\sum_{j=1}^n\ln {Q}(y_j\mid \bx)\big\}.	
\end{align*}
So the local terms $(\ln {P}(x_i))_{i=1}^p$ act as external magnetic fields, while the log-likelihood terms $(\ln {Q}(y_j\mid \bx))_{j=1}^n$ are interactions between spins that correlate them in a highly non-trivial way. If these were pairwise, we would recover an instance of the \textbf{Ising model}, see below.

Coming back to the notion of optimal estimators: with our statistical mechanics interpretation in mind, we now understand that MAP estimation is equivalent to \textbf{finding the ground state}, namely the spin configuration that minimizes the energy. Instead computing the MMSE estimator relies on \textbf{sampling the posterior/Gibbs-Boltzmann distribution}; these are among the main algorithmic tasks in statistical mechanics, and correspond exactly to the inference task.

Suddenly it becomes almost obvious that high-d inference and statistical mechanics are very close cousins. We will discuss concrete examples. Yet this re-interpretation is not simply an observation: it allows to import the massive amount of analysis techniques and concepts developed for more than a century in statistical mechanics into the world of inference and data-sciences.

\subsection{Paradigmatic models of statistical mechanics, and order parameters}

One physicist reflex that spreaded in virtually all scientific areas is to consider a toy model containing all the relevant features of more complex/realistic models, but yet being simple enough to be tackled analytically in order to understand fundamental phenomena that should apply more broadly. Let us start by introducing two such models in statistical physics. We will later connect them to inference.

By far the better understood and most paradigmatic model in statistical mechanics is the \textbf{fully-connected Ising model}, also called \textbf{Curie-Weiss model (CW)}, defined by the Hamiltonian ($J>0$, $h\in \mathbb{R}$ and $\boldsymbol{\sigma}\in\{-1,1\}^p$) 
\begin{align*}
\mathcal{H}_{\text{CW}}(\boldsymbol{\sigma};J,h)=-\frac{J}{p}\sum_{i<j}^p\sigma_i\sigma_j-h\sum_{i}^p\sigma_i.	
\end{align*}
Statistical mechanics uses macroscopic quantities called \textbf{order parameters} for describing a complex system. For the CW model it is simply the \textbf{magnetisation} $$m_p(\boldsymbol{\sigma}):=\frac1p\sum_{i=1}^p \sigma_i.$$ Then $m:=\lim_{p} \langle m_p(\boldsymbol{\sigma}) \rangle$ describes whether the system is in an ordered \textbf{ferromagnetic phase} (if $m\neq 0$) or a disordered \textbf{antiferromagnetic (or ergodic) phase} (if $m=0$)\footnote{This is true whenever $h\neq 0$. In a system with a global sign-flip symmetry like here when the external field is null $\mathcal{H}_{\text{CW}}(\boldsymbol{\sigma};J,h=0)=\mathcal{H}_{\text{CW}}(-\boldsymbol{\sigma};J,h=0)$, one needs to break this symmetry by introducing a small external field, and then taking the limit of this field to $0$ after the thermodynamic limit. The resulting magnetisation value $m^\pm:=\lim_{h\to 0^\pm}\lim_{p} \frac1p\sum_{i=1}^p\langle \sigma_i\rangle_h$ will depend, below its critical temperature $1/\beta_{\text{c}}$, on whether the limit $h\to 0$ is taken from below or above.}; here we introduced the standard physics notation $\langle\, \cdot\,\rangle$ to denote an average with respect to the Gibbs-Boltzmann distribution \eqref{GB}. In this simple model the concentration of measure implies $m_p(\boldsymbol{\sigma})=m+o_p(1)$.  

Another important model of spin system is the disordered version of the CW model. The \textbf{Sherrington-Kirkpatrick (SK) model}, or \textbf{mean-field spin glass}, is defined by the Hamiltonian
\begin{align}\label{SK_ham}
\mathcal{H}_{\text{SK}}(\boldsymbol{\sigma};\mathbf{J},h)=-\sum_{i<j}^p\frac{J_{ij}}{\sqrt{p}}\sigma_i\sigma_j-h\sum_{i}^p\sigma_i.	
\end{align}
The quenched interactions $J_{ij}\sim \mathcal{N}(0,1)$ are i.i.d. realisations of a normal random variable.

Let us open a technical parenthesis that is not crucial for the remaining of the discussion. A single scalar order parameter like the magnetisation is not enough anymore to describe the phenomenology of the SK model. Instead one needs to consider a richer distributional order parameter $$\mathbb{P}(q):=\lim_{p} \mathbb{E}_{\mathbf{J}}\mathbb{P}(q_p\mid \mathbf{J}),$$ which is the asymptotic probability distribution of the \textbf{overlap} $$q_p:= \frac 1p \sum_{i=1}^p \sigma_i^{(1)}\sigma_i^{(2)}.$$ Here $\boldsymbol{\sigma}^{(1)}$ and $\boldsymbol{\sigma}^{(2)}$ are (conditionally on $\mathbf{J}$) i.i.d. random vectors drawn from the same Gibbs-Boltzmann distribution; there are often called replicas. In the case of the CW model the asymptotic distribution of the magnetisation was simply a dirac mass $\delta_{m}$, so $m$ fully described the system. But here the concentration of measure is much more subtle: the overlap does not concentrate at low temperature ($q_p\neq \lim_{p}\mathbb{E}_{\mathbf{J}}\langle q_p\rangle+o_p(1)$), but its distribution does: $\mathbb{P}(q_p\mid \mathbf{J})$ converges in distribution to $\mathbb{P}(q)$ as $p\to \infty$. The shape of this distribution then allows to describe the various phases of the model (ferromagnetic, antiferromagnetic, spin glass, etc). Therefore the precense of disorder $\mathbf{J}$ changes drastically the model and its phenomenology, and describing it goes beyond the scope of this article, see \cite{mezard2009information,panchenko2013sherrington} for more details\footnote{The solution of the SK model has been found by G. Parisi \cite{parisi1980sequence,mezard1987spin} and the rigorous proof of the Parisi solution obtained by F. Guerra \cite{guerra2003broken} and M. Talagrand \cite{talagrand2006parisi} (and later re-proved by D. Panchenko \cite{panchenko2013sherrington}).}. The SK model has generated a whole field of research at the crossroad of physics, information theory, computer science and mathematics. And as we will realize soon, this model and its non-disordered cousin the CW model are both deeply connected to statistical inference too. 

What is the order parameter in high-d inference? A natural candidate is an error metric, and well will focus on the MMSE \eqref{MMSE}. The MMSE characterizes the \textbf{information-theoretic phases}: information-theoretically possible inference phase where the MMSE is relatively small, and the  impossible inference regime where it is comparatively high. Note that in general the location of the \textbf{information-theoretic phase transition} separating these phases does not depend on which error metric is used to probe the \textbf{phase diagram}; we will come back to these notions.

\subsection{Thermodynamics: free entropy and phase transitions}
A key quantity in statistical mechanics is the \textbf{free entropy} (or minus the \textbf{free energy}):
\begin{align*}
f_p=\frac{1}{\beta p}\ln \mathcal{Z}.
\end{align*}
It contains all thermodynamic information about the model: the non-analyticity points of its thermodynamic limit $f:=\lim_{p}f_p$ correspond to the location of \textbf{phase transitions}. A phase transition is when a complex system experiences a change in the behavior of certain order parameters when external \textbf{control parameters} are varied. The canonical example is water, whose phase may by characterized by a local density of molecules or an average correlation lenght (two order parameters) while the temperature and/or the pressure evolve (two control parameters).

One of the main use of the free entropy is that it allows to access (some) order parameters and their fluctuations; it is the moment generating function, the order parameter(s) being the moments. E.g., in the CW model the magnetisation and its flucutuations are obtained by taking derivatives w.r.t. the field $h$:
\begin{align}
f_p'	=\langle m_p\rangle, \qquad  f_p''= p\langle (m_p-\langle m_p\rangle)^2\rangle\label{fprime}
\end{align}
where the symbol $'$ means a $h$-derivative. For disordered systems like the SK model, we generally consider the expected free entropy 
\begin{align*}
\mathbb{E}f_p=\frac{1}{\beta p}\mathbb{E}_{\mathbf{J}}\ln \mathcal{Z}(\mathbf{J}).	
\end{align*}
It is equivalent to the non-averaged free entropy, but more practical to compute as independent of a particular realisation of the interactions $\mathbf{J}$. The equivalence is again a consequence of the concentration of measure, that implies $$f_p(\mathbf{J})=\lim_{p}\mathbb{E}f_p+o_p(1).$$ Note that even if the overlap is not self-averaging, like in the SK model and other spin glasses at low temperature (or combinatorial optimisation problems \cite{mezard2009information}), the free energy is always self-averaging (for well defined models).

Let us discuss further the notion of phase transition. There exist many types of phase transition; sometimes they are quite smooth (these are of the \textbf{second order} type because they correspond to a discontinuity of a second-order derivative of the asymptotic free entropy), and sometimes very sharp and discontinuous (of the \textbf{first order} type, namely, a discontinuity of a first order derivative of $f$).  Examples of phase transitions are: the recovery of a souvenir by the brain once enough stimuli in the direction of the memorized pattern are provided (a simple model of associative memory is the \textbf{Hopfield model}). Here the order parameter is the overlap with the memorized pattern and the control parameter is the amount of stimuli. A crack in the financial market, where suddenly all prices drop all together. The sudden rigidity transition that happens when you randomly pack enough balls in a box (this is called the \textbf{jamming transition}, and this is related to computer memory optimization or error correcting codes in communication). When communicating bits through a given noisy channel, there exists a maximum rate of information transmission; communication above this sharp threshold is impossible as information gets lost due to noise. This limit is called the \textbf{Shannon capacity} \cite{shannon1948mathematical,cover1999elements} and really is nothing but a phase transition. The order parameter is the quality of recovery of the information bits, the control parameter is the communication rate. A final one. Say you want to train a classification algorithm that, when given a large data-base of labeled training examples, is able to distinguish pictures of dogs and cats. There exists a minimum number of training examples below which, no matter the power of the computer, the algorithm will never be able to properly classify the images; this is an information-theoretic transition. The order parameter is the classification performance of the algorithm, the control parameter is the size of the training set, see, e.g., \cite{barbier2019optimal,youtube}

\subsection{Information theory: Shannon entropy and mutual information}

Information theory in the context of inference is mainly concerned with the following question: when does data contains enough information so that it can be used to infer something about the process that generated it?

To adress this question and, connected to that, understand what is the cousin notion of free entropy in high-d inference, we first need to recall what is the \textbf{Shannon entropy}. The understanding of this object is fundamental and gave rise to the birth of information theory \cite{shannon1948mathematical}, so we will spend some time to discuss it in details. As understood by Shannon, it is related to the older notion of entropy in thermodynamics and statistical mechanics as we will see, thus the name. Its definition for a discrete r.v is
\begin{align*}
H(\bx):=\sum_{\bx}\mathbb{P}({\bx})\ln\frac{1}{\mathbb{P}({\bx})}=\mathbb{E}_{{\bx}}\ln\frac{1}{\mathbb{P}({\bx})}.	
\end{align*}
We slightly abuse notation and use the same symbol $\bx$ for a r.v. and its outcome (while information theory usually denotes the r.v. in capital letter $\mathbf{X}$ and an outcome $\bx$). We will restrict our discussion to discrete r.vs. as the interpretation is a bit more subtle in the continuous case, but a lot of the intuition generalizes. 

The Shannon entropy $H(\bx)=\mathbb{E}_\bx h(\bx)$ is the expectation of the \textbf{information content} $$h(\bx):=-\ln \mathbb{P}(\bx),$$ or \textbf{surprise}, of the outcome $\bx$. If this outcome has low probability then observing it is quite surprising, and it brings a lot of information as it was not expected: $h(\bx)$ is high. If instead $\mathbb{P}(\bx)$ is close to $1$ it is not surprising to observe $\bx$, so this outcome brings low information: $h(\bx)$ is low. Said differently: if the outcome of a r.v. is very probable, it is no surprise (and generally uninteresting) when it happens, because it was expected. However, if an outcome is unlikely to occur, it is much more informative if it happens to be observed. The term information content must be understood as a \emph{potential} information gain if $\bx$ is observed. When using the $\log_2$ the information content and entropy are expressed in ``bits''.

Imagine you are in the desert and suddenly it rains like hell. Worst, it rains cows that play piano! What?! It is amazingly surprising no? The probability of this event is actually so low that it brings an enormous amount of information; in this case it should lead you to the conclusion that you are dreaming. If instead your are in the desert and its super sunny and hot, it is not surprising at all; this does not bring more information than what you already know, and if you are dreaming, it is unlikely that this observation will help you realize it. Another example: the knowledge that some particular number will not be the winning one of a lottery provides very little information, because any particular chosen number will almost certainly not win. However, knowledge that a particular number will win a lottery has high informational value because it communicates the outcome of a very low probability event. 

The entropy can also be interpreted as a \emph{measure of unpredictability} of the r.v. $\bx$, or of \emph{uninformation/lack of knowledge} about what $\bx$'s outcome will be: the more surprising are the outcomes in expectation, the more unpredictable is the actual outcome, which also mean the less we know about $\bx$ \emph{before} observing it. $H(\bx)$ \emph{quantifies the expected amount of missing information necessary to determine the outcome of $\bx$ before observing it}. This can be confusing because previously we said that $H(\bx)$ is an expected information content, while now we speak about a measure of uninformation. There is no paradox: an information $H(\bx)$ is \emph{gained} in average when $\bx$'s outcome \emph{is actually observed}. But \emph{prior} to observing the outcome, $H(\bx)$ is a measure of uninformation about it. Put differently: observing the outcome $\bx$ \emph{converts} in average $H(\bx)$ units of uninformation into information. So it just a matter of conceptually placing ourselves \emph{before} $\bx$ is observed --in which case the interpretation as a measure of uninformation may be more natural--, or \emph{after} $\bx$ is observed --where the interpretation as an expected information content seems to fit better. But at the end this is the same thing. 

An example might help: the outcome of a toss of a fair coin $x_{\text{fair}}\sim\text{Ber}(1/2)$ is much more unpredictable than the outcome of a strongly biased coin $x_{\text{bias}}\sim\text{Ber}(9/10)$, or equivalently our lack of knowledge about what will be $x_{{\text{fair}}}$ is higher: we are more uninformed. But when \emph{observing} the outcome of the fair coin, we then \emph{gain} more information than with the unfair one, because it is in average more surprising. In the first case, which has entropy $H(x_{\text{fair}})$ of one bit, betting on one side or the other is the same statistically. While in the second case, where $H(x_{\text{bias}})=\frac{9}{10}\log_2\frac{10}{9}+\frac1{10}\log_2 10\approx 0.47$, the outcome is much more predictable, we are less uninformed (= more informed); it would be an error not to bet on the outcome $x_{\text{bias}}=1$. 

To summarize: the Shannon entropy $H(\bx)$ of the r.v. $\bx$ quantifies: $i)$ its average information content, i.e., the expected information gain when observing outcome $\bx$; $ii)$ the average uninformation/lack of knowledge about the outcome $\bx$ prior to observe it; $iii)$ its unpredictability. The higher the entropy of $\bx$, the less ``structured'' its distribution is; $v)$ when expressed in bits $H(\bx)$ is the expected number of binary ``yes/no'' questions required to determine the outcome \emph{before} it is observed, or equivalently, the expected number of binary questions that the oucome $\bx$ has answered \emph{after} being observed.

Similarly the \textbf{conditional entropy} is:
\begin{align*}
H(\bx\mid \by):=\sum_{\bx,\by}\mathbb{P}({\by})\mathbb{P}(\bx\mid {\by})\ln\frac{1}{\mathbb{P}({\bx}\mid \by)}.	
\end{align*}
It is the expected information revealed by evaluating the outcome of $\bx$ given that you know already the outcome of $\by$. Or equivalently, it is the expected remaining amount of unpredictability of $\bx$ given that $\by$ has already been observed.

The entropy has many important properties that make it a ``good'' definition of information content, one of the main being that it is additive for independent r.vs.: $$H(\bx,\by)=H(\by)+H(\bx) \quad \text{if}  \quad \mathbb{P}(\bx,\by)=\mathbb{P}(\bx)\mathbb{P}(\by),$$ and many other ones such as its non-negativity (for discrete r.vs.) and the chain rule $$H(\bx,\by) = H(\bx\mid \by) + H(\by) = H(\by\mid \bx) + H(\bx).$$ (It is easy to prove these facts from the definition of the entropy). But all these justifications are not enough to \emph{prove} that it is indeed \emph{the} correct definition. Maybe other functions verify all these properties and have a similar interpretation. The mathematical proof that the entropy indeed is the correct definition comes from the \textbf{source coding theorem} of C.~Shannon, the father of information theory, see \cite{shannon1948mathematical,mackay,cover1999elements}. Let us describe it in words. We consider binary symbols but the following reasonning applies to more generic discrete alphabets. 

Roughly, the source coding theorem says that if a source generates strings $(x_1,x_2,\ldots,x_n)$ of $n\gg 1$ binary symbols that are i.i.d. outcomes of some random variable $x$, then there exists a compressed \textbf{code} $\mathcal{C}_\delta$ for this source of cardinal $|\mathcal{C}_\delta|\approx 2^{n H(x)}\le 2^n$, and this independently of the risk $0<\delta<1$ we are ready to take in losing information when coding (as $n\to\infty$). 

Let us understand what does that mean, and why it implies that $H(x)$ is the proper definition of information content carried by the random variable $x$. 
\begin{enumerate}[label=\alph*)]
\item First, why introducing a source of long strings? The information content of the source, \emph{whatever it means}, must be $n$ times the one of $x$ alone because information must be additive for independent variables $(x_1,x_2,\ldots, x_n)$. As a consequence the expected information content per symbol of the source equals the one of $x$. So studying the source or $x$ is the same from an information-theoretic point of view. But as $n$ will get large, Shannon understood that the concentration of measure in the form of the law of large numbers will help a lot in the analysis.
\item What is a code? A code $\mathcal{C}_\delta$ is any alternative ``maximally compressed'' represention of the set of strings. Namely it is a set of smaller cardinal than $2^n$, the number of possible strings, such that a random string from the source has an associated element in the code with probability (over the strings) at least $1-\delta$. And at the same time the code has smallest possibe cardinal. A code $\mathcal{C}_\delta$ therefore ``encodes'' part of the information (as tuned by $\delta$) about the source through a bijective mapping between $\mathcal{C}_\delta$ and a subset of the $2^n$ possible strings. The risk we take is in the sense that with probability $<\delta$ a string generated by the source will not have an associated element in $\mathcal{C}_\delta$ so its information is lost when coding. A constructive way of defining $\mathcal{C}_\delta$ is to rank all possible strings according to their probability. Then add a first element $C_1$ in $\mathcal{C}_\delta$, associated to the most probable string. Then add a second element $C_2$ in $\mathcal{C}_\delta$ mapped to the second most probable string, and so on, until the sum of probabilities of the strings mapped to the code elements exceeds $1-\delta$.
\item The information content of this code expressed in bits is naturally defined as the number of binary symbols necessary to represent any element of the code: $\log_2|\mathcal{C}_\delta|$. Moreover Shannon showed that $\log_2|\mathcal{C}_\delta| \to n H(x)$ as $n\to \infty$. \emph{A-priori} this information content is less than the one of the original source as some risk $\delta$ has been taken when compressing/mapping the source to $\mathcal{C}_\delta$; we talk about \textbf{lossy compression}.
\item But the crucial observation is that the quantity $H(x)$ defining the cardinal of the code necessary to compress the source up to risk $0<\delta<1$ becomes \emph{independent} of the risk in the limit $n\gg 1$. This means that as long as we allow ourselves a tiny probability of error $\delta$ (independent of $n$), compression down to $nH(x)$ bits is possible. But even if we are allowed a large probability of error, we still can compress the source only down to $nH(x)$ bits. This strongly suggests that $nH(x)$ is the fundamental information content of the source. As a consequence of this and point ${\text a})$ the information content of $x$ is also $\frac1n \log_2|\mathcal{C}_\delta|\to H(x)$. This ends the reasonning. 
\end{enumerate}

Let us give a high-level proof of the source coding theorem. The point is that, as $n$ gets larger, by the law of large numbers almost all strings actually generated by the random source are \emph{typical}, so that only the typical sequences need to be encoded during the compression (the others are too unprobable and therefore are not coded). Let us consider for simplicity Bernoulli variables $x\sim \text{Ber}(\rho)$. All typical sequences have approximately the same number $n\rho$ of ones and $n(1-\rho)$ of zeros. Indeed, the probability that the string has exactly $R$ ones is a binomial distribution $R\sim \text{Bin}(n,\rho)$.
The relative fluctuation of $R$ is $O(1/\sqrt{n})$ so $R$ concentrates onto its mean when $n$ gets large\footnote{Relative fluctuations of the order $O(1/\sqrt{n})$ of macroscopic quantities like $R$ are typical of complex systems treated in statistical mechanics. That the relative fluctuations vanish is the reason why such random systems can be analyzed and described by asymptotically (as $n\to+\infty$) deterministic observables, converging on their ensemble mean.}. This implies that with high probability the only possibly observed strings are those with $R$ values very close to $n\rho$: this informally defines the \emph{typical set}. So the probability of a typical sequence $\bx_{\text{typ}}=(x_1,\ldots,x_n)$ is 
\begin{align*}
\mathbb{P}(\bx_{\text{typ}}) = \prod_{i}^n P(x_{\text{typ},i})\approx  \rho^{n\rho}(1-\rho)^{n(1-\rho)} .\
\end{align*}
Denote this probability of a typical string $P_{\text{typ}}:=\rho^{n\rho} (1-\rho)^{n(1-\rho)}$.
What is the information content/surprise in bits of a typical outcome? 
\begin{align}
    \log_2\frac{1}{P_\text{{typ}}} &= -n(\rho\log_2\rho+(1-\rho)\log_2(1-\rho)) = nH(x). \label{probaTyp}
\end{align}
So the proof strategy is: $i)$ as $n$ gets large only typical sequences/outcomes are observed; they carry almost all the probability mass. So when defining the code $\mathcal{C}_{\delta}$ we need only to code these typical outcomes; doing so it is maximally compressed. The number of typical strings is exponentially large in $n$ (this follows from the \textbf{asymptotic equipartition principle}), so even if we allow a risk $\delta$ very close to $1$ (but independent of $n$) and therefore only code a small fraction of the typical sequences, there are still approximately as many at leading (exponential) order as $n$ gets large. E.g., if there are $\exp(an)$ typical sequences and we only code $(1-\delta) \exp(an)=\exp(an+\ln(1-\delta))$ of them, there are the same number at leading order for any fixed $a>0$ and $1>\delta>0$ as $n\gg 1$. So independently of $\delta$ the number $|\mathcal{C}_{\delta}|$ of typical sequences necessary to code is the same at leading exponential order. $ii)$ The question then becomes: can we count them, i.e., evaluate $|\mathcal{C}_{\delta}|$ at leading order? By definition all typical sequences have approximately the same probability $P_{\text{typ}}$, and they carry almost all the mass. Therefore 
\begin{align*}
\sum_{\{\bx \, \text{typical}\}} \mathbb{P}({\bx})\approx \#_{\text{typ}} P_{ \text{typ}}\approx 1,	
\end{align*}
where $\#_{\text{typ}}$ is the number of typical sequences. With what we said previously $\#_{\text{typ}}$ equals $|\mathcal{C}_{\delta}|$ at leading order. This implies that there are approximately $\#_{\text{typ}}\approx 1/P_{\text{typ}}= 2^{nH(X)}$ typical sequences (from \eqref{probaTyp}). We can thus count them at leading order. This allows to estimate the expected information content per bit as $\frac1n \log_2|\mathcal{C}_{\delta}|\approx H(x)$, which is the same as the expected information content of $x$ by definition of the source. The same argument extends to more general (non binary) alphabet.

A connected information-theoretic quantity is the \textbf{mutual information}: 
\begin{align*}
I(\bx;\by):=H(\bx)-H(\bx\mid \by)=H(\by)-H(\by\mid \bx).
\end{align*}
It is interpreted as a measure of the mutual dependence of $\bx$ and $\by$. It quantifies the ``amount of information'' obtained about one r.v. through observing the other one. And indeed it cancels if and only if the r.vs. are independent: 
$$I(\bx;\by)\ge 0 \  \text{with equality if and only if} \ \mathbb{P}(\bx,\by)=\mathbb{P}(\bx)\mathbb{P}(\by).$$ 

In an inference problem where we want to recover the parameters $\bx$ from the data $\by(\bx)$ the last form has a particularly nice interpretation: $H(\by)-H(\by\mid \bx)$ is the total information carried by the data minus the remaining unpredictability/uninformation about the data when the signal is known, which is therefore the noise contribution. E.g., in a Gaussian \textbf{denoising model} $$y=\sqrt{\lambda}\,x+z$$ with $z\sim\mathcal{N}(0,1)$ we have $H(y\mid x)=H(z)=\ln(2\pi e)/2$. The mutual information is thus the information carried by the data purely about the signal. As such it quantifies the information-theoretic limits of inference, and computing it in high-d settings is a key goal of information theory. In the Gaussian denoising model it is an exercise to show that its explicit expression reads (here $x^*,x$ are i.i.d. from $\mathbb{P}(x)$)
\begin{align}
  I(x&;y)\!=\!\frac{\lambda}{2}\mathbb{E}[x^2]\!-\!\EE_{x^*}\!\ln \mathbb{E}_x \exp\Big\{\lambda x^*x+\sqrt{\lambda}zx-\frac{\lambda}{2}x^2\Big\}.\label{MI_gauss}
\end{align}

\subsection{Denoising, the I-MMSE formula and the information-theoretic interpretation of the free entropy}

Consider the general denoising model, where $\bx$ can be a vector, matrix, etc: $$\by=\sqrt{\lambda}\,\bx+\mathbf{z}.$$ The r.v. $\mathbf{z}$ has same dimension as the signal and has i.i.d. standard normal entries. The SNR $\lambda>0$ controls the signal strenght: the higher, the easier is the inference task of recovering $\bx$ from $\by$. 

There exists a general identity called \textbf{I-MMSE formula} \cite{guo2005mutual} relating the mutual information and the MMSE for the denoising model:
\begin{align*}
\frac{d}{d\lambda}\frac1pI(\bx;\by)\!=\!\frac12\mathbb{E}_{\bx,\by}\text{MMSE}_p\!=\!\frac12\mathbb{E}\|\bx-\mathbb{E}[\bx\mid\by]\|^2.
\end{align*}
This relation is the equivalent of the thermodynamic identity $f_p'=\langle m_p\rangle$ in \eqref{fprime}, but for high-d inference (under Gaussian noise). 

Let us clarify the connection between free entropy and mutual information. The expected log-partition function in the Bayes formula reads 
\begin{align*}
\mathbb{E}_{\by} \ln \mathcal{Z}(\by)=\int d\by  \mathbb{P}(\by) \ln \mathbb{P}(\by)=-H(\by).	
\end{align*}
Therefore the expected free entropy is linked to the Shannon entropy of the data: 
\begin{align*}
-\mathbb{E}_\by f_p(\by)=\frac1pH(\by).	
\end{align*}
So the mutual information verifies
\begin{align}
\frac1pI(\bx;\by)=-\mathbb{E} f_p(\by)-\frac1pH(\by\mid \bx).	\label{11}
\end{align}
The term $\frac1p H(\by\mid \bx)=\frac1p H(\mathbf{z})=\frac12\ln(2\pi e)$ is trivial (because the noise has i.i.d. components). 

Another way to see the connection is by starting from the thermodynamic definition of free entropy: the Shannon entropy of the Gibbs-Boltzmann distribution (the posterior) minus the internal energy (recall $\beta=1$): 
\begin{align}\label{freeEnt_toUse}
p\mathbb{E} f_p(\by)=H(\bx\mid \by)-\mathbb{E}\langle \mathcal{H}(\bx;\by)\rangle,	
\end{align}
where $\mathcal{H}(\bx;\by)=-\ln \mathbb{P}(\by\mid\bx)-\ln \mathbb{P}(\bx)$ is the Hamiltonian defining the posterior. We focus on the Gaussian denoising model. Using the Bayes formula \eqref{Bayes} the internal energy verifies
\begin{align*}
&\int d\bx d\by \,\mathbb{P}(\by)\mathbb{P}(\bx\mid \by)  \mathcal{H}(\bx;\by)=\int d\bx d\by\, \mathbb{P}(\bx)\mathbb{P}(\by\mid \bx)  \mathcal{H}(\bx;\by)=\int d\bx d\bz\, \mathbb{P}(\bx)\mathbb{P}(\bz)  \mathcal{H}(\bx;\sqrt{\lambda}\,\bx +\bz)	
\end{align*}
using the change of variable $\by=\sqrt{\lambda}\,\bx+\mathbf{z}$. As the noise is i.i.d. Gaussian the likelihood $\mathbb{P}(\by\mid\bx)$ is a multivariate Gaussian measure with mean $\sqrt{\lambda}\,\bx$ and identity covariance, and thus $\mathbb{P}(\bz)$ is a standard multivariate Gaussian after the change of variable. Therefore
\begin{align*}
\mathcal{H}(\bx;\by)=\frac12\|\sqrt{\lambda}\,\bx-\by\|^2+\frac p2\ln(2\pi)-\ln \mathbb{P}(\bx)	.
\end{align*}
We finally reach that the internal energy
\begin{align*}
\mathbb{E}\langle \mathcal{H}(\bx;\by)\rangle&=\frac12\mathbb{E}\|\bz\|^2+\frac p2\ln(2\pi)-\mathbb{E}\ln \mathbb{P}(\bx)=\frac p2\ln(2\pi e)+H(\bx).	
\end{align*}
Using this as well as \eqref{freeEnt_toUse} in $\frac1pI(\bx;\by)=\frac1pH(\bx)-\frac1pH(\bx\mid \by)$ we recover \eqref{11} and $-\mathbb{E} f_p=\frac1pH(\by)$. Therefore a physicist trying to compute the free entropy and an information theorist the mutual information are actually aiming for the very same goal. 

Thanks to the I-MMSE relation the MMSE order parameter can be derived from $I(\bx;\by)$, or the magnetisation from the free entropy in statistical mechanics models, at least ``in theory''. Indeed, computing the $p$-dimensional integrals necessary to obtain the thermodynamic potentials (mutual information, free entropy) or the order parameters ``directly'' is generally a daunting task. But as we will discuss towards the end, in some high-d problems, this can be reduced to a (much) simpler scalar optimisation problem thanks to the concentration of measure phenomenon.

Consider a factorized prior $\mathbb{P}(\bx)=\prod_{i}P(x_i)$. In this setting, is the denoising model a ``good'' example of high-d inference problem, with phase transitions and a rich phase diagram? No. Indeed in model $\by=\sqrt{\lambda}\,\bx+\mathbf{z}$ each data point $y_i(x_i,z_i)$ is only function of a single signal and noise components. The r.vs. $(x_i,z_i)_{i\le p}$ are i.i.d. by the factorization assumption. As a consequence the MMSE of the whole signal $\bx$ equals the MMSE of a single entry as they are all statistically equivalent: $\mathbb{E}\,\text{MMSE}_p=\mathbb{E}[(\mathbb{E}[x_1\mid {y}_1]-{x_1})^2]$. This quantity is easily shown to be 
\begin{align}
\mathbb{E}[(\mathbb{E}[x_1\mid {y}_1]-{x_1})^2]=\mathbb{E}[x^2]-\mathbb{E}_{z,x^*}\Big[x^*\frac{\mathbb{E}_{x}\, x\,e^{-\frac12(\sqrt{\lambda}(x^*-x)+z)^2} }{\mathbb{E}_{x}\, e^{-\frac12(\sqrt{\lambda}(x^*-x)+z)^2}}\Big]\label{MMSE_denoi}	
\end{align}
where $x,x^*$ are i.i.d. from $P$ and $z$ is a standard Gaussian r.v. Plotting this MMSE order parameter as a function of the $\lambda$ control parameter, we get a smooth continuous non-increasing curve, that vanishes as $\lambda\to \infty$. Not so exciting. This is because the variables $(x_i)$ are in fact \textbf{decoupled} and the problem collapses onto $p$ parallel equivalent low-dimensional/scalar inference problems $y_i=\sqrt{\lambda}\, x_i+z_i$. And all are statistically equivalent so studying one is enough. Something is missing in the model in order to turn the picture into something richer. The denoising model lacks a key ingredient of complex systems: \textbf{correlations} among the signal entries induced by non-trivial interactions between the $(x_i)$ in the Hamiltonian.

\section{A paradigm of high-d inference: the spike Wigner model}

In high-d inference an important model is the \textbf{spike Wigner (SW) model}, also called \textbf{low-rank matrix factorisation}. As we will discuss it is a close cousin of the Ising and SK models in statistical mechanics. It was introduced in random matrix theory as a simple model of principal components analaysis \cite{WSM}, which is the most widely used dimensionality reduction technique.

Let $\mathbf{z}=(z_{ij})_{i,j=1}^n$ be a noise matrix with independent i.i.d. standard normal entries $z_{ij}\sim\mathcal{N}(0,1)$; this is called a Wigner matrix. In the SW model the data is (the upper triangular part of) $\mathbb{R}^{p\times p}\ni \by= \sqrt{\lambda/p}\,\bx\bx^\intercal + \mathbf{z}$, or componentwise,
\begin{align}
y_{ij}=\sqrt{\frac{\lambda}{p}}x_ix_j+z_{ij} \quad \mbox{for} \quad  1\le i<j\le p. \label{WSM}
\end{align}
The signal $\bx$ is a realisation of the prior $\mathbb{P}(\bx)=\prod_{i=1}^p P(x_i)$. Using that the likelihood is the standard multivariate Gaussian measure the posterior reads (constant terms are simplified with the normalization)
\begin{align}\label{postSW}
\mathbb{P}(\bx\mid\by)=&\frac1{\mathcal{Z}(\by)}\exp\Big\{\sum_{i=1}^p\ln P(x_i)-\frac12 \sum_{i<j}^p(y_{ij}-\sqrt{\frac\lambda p}\,x_ix_j)^2\Big\}.
\end{align} 
Now we see pairwise interactions in the Hamiltonian, so the $(x_i)$ are not anymore decoupled and the model cannot be reduced to independent scalar inference problems: this system \emph{is} complex.

Note that the information about the sign of $\bx$ is lost by $\pm \bx$ symmetry in this measure whenever $P(x_i)$ is even, e.g., when considering a signal with $\pm1$ uniform entries. In such situations $\mathbb{P}(\bx\mid\by)=\mathbb{P}(-\bx\mid\by)$ so that $\mathbb{E}[\bx\mid\by]=(0)$. Therefore it makes more sense in general to consider the rank-one matrix $\bx\bx^\intercal=(x_ix_j)_{i,j=1}^p$ as hidden signal (called ``spike''). Anyway if the statistician can recover the spike, it may access $|\bx|$ by finding its eigenvector. The noise $\mathbf{z}$ represents a uncontrolled source of randomness that corrupts the spike. The statistician task is then to infer $\bx\bx^\intercal$ as accurately as possible given $\by$ and the knowledge of the data-generating process (namely the model \eqref{WSM}, but not the specific realization of $\bx$ nor $\mathbf{z}$). We could generalise to other type of noise (not only Gaussian nor additive), but the qualitative picture would not change much.

The scaling $1/\sqrt{p}$ of the SNR in \eqref{WSM} is there to make the inference task nor impossible nor trivial. Any other scaling would turn the problem, in the large-system limit $p\to \infty$, into a model with not much interest. By ``uninteresting'' we mean that the (asymptotic average) spike-MMSE 
\begin{align*}
\text{MMSE}:=\lim_{p}\frac1{p^2} \mathbb{E}\|\mathbb{E}[\bx\bx^\intercal\mid \by]-\bx\bx^\intercal\|^2	
\end{align*}
would be essentially equal to $0$ for a scaling $p^{-\gamma}\gg 1/\sqrt{p}$, or to its maximum value for a scaling $p^{-\gamma}\ll 1/\sqrt{p}$, and this independently of $\lambda=O(1)$. Here $\|\, \cdot\,\|$ is the Frobenius norm and $$\mathbb{E}[\bx\bx^\intercal\mid \by]:=\int d\bx\,\bx\bx^\intercal \,\mathbb{P}(\bx\mid \by)$$ is the MMSE estimator of the spike. But precisely for the scaling $\gamma=1/2$ a rich \textbf{phase diagram} emerges with \textbf{information-theoretic phase transitions}.

Let us understand more precisely why this is the proper SNR scaling, and connected to that, that we are indeed in the high-d regime. 
%
%
For the inference task not to be trivial we need place ourselves in the high-d regime. As we explained already this means that the total SNR per parameters, i.e., $\!\# \,\text{data points} \times \text{SNR}_{\text{d}} \div \# \,\text{parameters\ to\ infer}$, should tend to an order $1$ constant in the thermodynamic limit. We have access to $p(p-1)/2$ conditionally independent data points $(y_{ij})_{i<j}$ and $\text{SNR}_{\text{d}}=\mathbb{E}[(\sqrt{\lambda/p}\,x_ix_j)^2] =(\mathbb{E}[x_1^2])^2 \lambda/p$. 
%
%
We verify that
\begin{align*}
(\mathbb{E}[x_1^2])^2\Big(\frac p2(p-1)\times\frac\lambda p\Big)\frac1p =(\mathbb{E}[x_1^2])^2\frac\lambda 2+O\Big(\frac1p\Big)
\end{align*}
is indeed $O(1)$ as we assume $(\mathbb{E}[x_1])^2=O(1)$. This explains the scaling $1/\sqrt{p}$ in the observation model \eqref{WSM}: we are in the high-d regime.

Examples of applications of this model are (we consider in all cases that the prior factorizes as $\mathbb{P}(\bx)=\prod_{i\le p}P(x_i)$):
\begin{itemize}
  \item \textbf{Sparse principal components analysis:} In the simplest case the prior $P= \text{Ber}(\rho)$ is Bernoulli. The task is to estimate the hidden sparse low-rank representation $\bx \bx^\intercal$ of ${\by}$. 
  \item \textbf{Submatrix localization:} Again $P = \text{Ber}(\rho)$. One has then to extract a submatrix of ${\by}$ of size $\rho p\times \rho p$ with larger mean than the background noise matrix; this is an important model of hidden structure in computer science.
  \item \textbf{Community detection in the stochastic block model (SBM):} The (assortative) SBM is a network model where edges between nodes belonging to the same community are more probably observed. Given these observed edges, the task is to infer the community to which belong each nodes. For example, assume you know the network of friendships in some social network. Under the hypothesis that people voting for the same political party (among two) are connected in this network with higher probability than when they vote opposite parties, is it possible to guess the two communities of voters (up to a global permutation)?

  Recovering two communities of size $\rho p$ and $(1-\rho)p$ in a SBM of $p$ vertices is information-theoretically ``equivalent'' to the SW model with prior (see \cite{lelarge2019fundamental} for the precise meaning of equivalence)
\begin{align}
   P =\rho \delta_{\sqrt{(1-\rho)/\rho}}+(1-\rho)\delta_{-\sqrt{\rho/(1-\rho)}}.\label{SBM}
\end{align}
	
  \item $\mathbb{Z}/2$\textbf{-synchronization:} The prior is Rademacher $P= \frac12\delta_{-1} + \frac12\delta_1$. The task is to infer the nodes states $\bx\in\{-1,1\}^p$ (up to a global sign) from noisy pairwise products $\by$. 

  A possible interpretation: imagine that you can ask to pairs $(i,j)$ of individuals whether they agree (+1) or not (-1) on some binary ``yes/no'' question, but you cannot ask to any individual $i$ alone what is her/his opinion on the question, and you have no a-priori idea about it. Moreover the answers $(y_{ij})$ you collect are transmitted through a very noisy (Gaussian) communication channel. Naively, you would naturally guess that the pair of individuals $(i,j)$ have the same opinion whenever $y_{ij}$ (equal to $\sqrt{\lambda/p}\,x_ix_j+z_{ij}$, where $x_i$ is the opinion of individual $i$) is positive because $z_{ij}$ is centered. For the pairs such that $y_{kl}<0$ you would bet instead that $x_kx_l=-1$ (i.e., that they disagree). Of course with this naive approach contradictions will appear because of the noise. Let us say that you collected such noisy answers $(y_{ij})$ for many (all) pairs. Can you optimally infer the opinion of each individuals (up to global flip), i.e., who are the ``synchronized individuals''? The naive approach is sub-optimal. What one needs to do is to use the posterior \eqref{postSW} and compute the MMSE estimator (in case the goal is to minimize the MSE) or the MAP estimator (if instead one wants to maximize the probability of finding $\bx$).
\end{itemize}

\subsection{Link to the Curie-Weiss and Sherrington-Kirkpatrick models}
As promised we now establish a clear connection between the CW and SK models from statistical mechanics and the SW model from high-d inference.

We consider the binary case $\bx\in\{-1,1\}^p$ with Rademacher prior. This corresponds to the $\mathbb{Z}/2$-synchronization problem discussed above. In this section it will be convenient to make appear at the same time both the ground-truth signal and the variable that is distributed according to the posterior. Therefore we will rename the ground-truth signal $\bx^*$ where the $*$ emphasizes that it is the true one, that is fixed when performing inference, while $\bx$ are the variables/spins that fluctuate according to the posterior. In the Rademacher case the prior gives a constant contribution that simplifies with the partition function and can therefore be dropped in the posterior. Then, as seen from \eqref{postSW}, the Hamiltonian of the SW model reads, when expressing the data as a function of the signal and noise using $y_{ij}= \sqrt{\lambda/p}\,x_i^*x_j^* + z_{ij}$ and simplifying all $\bx$-independent terms with the normalization,
\begin{align*}
	\mathcal{H}_{\text{SW}}(\bx;\by)= -\sum_{i<j}^p \Big( {\frac{\lambda}{p}}x_i^*x_j^*+\sqrt{\frac{\lambda}{p}}z_{ij}\Big)x_ix_j
\end{align*}
with $\bx\in\{-1,1\}^p$. This is exactly the SK Hamiltonian \eqref{SK_ham} when only the noise term $z_{ij}$ is present and $\lambda$ is set to one. The additional signal-related term $-\sum_{i<j}{\frac{\lambda}{p}}x_i^*x_j^*x_ix_j$ is called \textbf{planted term}, and inference models are \textbf{planted statistical mechanics models}. The planted term plays the role of external magnetic field that tends to align the spins in the signal direction; it carries the information. In contrast the noise term, that competes with the planted one, tends to align the spins in a random direction that is uncorrelated with the signal. Depending on the value of the SNR $\lambda$ that plays a similar role as the inverse temperature $\beta$, one term wins against the other: for high enough $\lambda>\lambda_{c}$ the planted term wins and the spins ``magnetise/polarise'' in the signal direction. Here $\lambda_{c}$ is the so-called \textbf{information-theoretic threshold} (see next section for more details). This polarisation is quantified by the overlap between a sample $\bx$ from the posterior and the signal $\bx^*$ 
\begin{align}
m_p^*=\frac1p \sum_{i=1}^px_ix_i^*	.\label{planted_mag}
\end{align}
Let $m^*:=\lim_p \mathbb{E}\langle m_p^*\rangle$. We write the posterior mean $\mathbb{E}[\,\cdot \mid \by]$ using the bracket notation $\langle \, \cdot\,\rangle$ from statistical mechanics to emphasize that $\mathbb{P}(\bx\mid\by)$ is a Gibbs-Boltzmann distribution; $\mathbb{E}$ is the average over all quenched variables $(\bx^*,\by)$ (or equivalently $(\bx^*,\bz)$). After some manipulations one can demonstrate that the expected spike-MMSE relates to this overlap order parameter as 
\begin{align*}
\mathbb{E}\,\text{MMSE}_p&=\frac1{p^2} \mathbb{E}\|\bx^*(\bx^*)^\intercal-\langle \bx\bx^\intercal\rangle \|^2=(\mathbb{E}[(x^*_1)^2])^2-\mathbb{E}\langle(m_p^*)^2\rangle.
\end{align*}
As in the CW model the concentration of measure implies (in the Bayesian optimal setting): $m_p^*=m^*+ o_p(1)$, and therefore concentration of the expected MMSE (and actually of the non-averaged one as well) towards the asymptotic average MMSE as $p\to\infty$: 
\begin{align}\label{MMSE_ptoMMSE}
\text{MMSE}_p\to(\mathbb{E}[(x_1^*)^2])^2-(m^*)^2\!=:\!\text{MMSE}.
\end{align}

Despite the Hamiltonian $\mathcal{H}_{\text{SW}}$ ressembles a lot the one of the SK as there is disorder, the phenomenology of the SW model is closer to the one of the CW model due to the planted term. The order parameter $m_p^*$ is the counterpart in planted problems of the magnetisation $m_p$ in the CW model, and it concentrates, while the overlap does not in the SK model; that makes a huge difference. This complicates drastically the analysis of the SK model, see \cite{panchenko2013sherrington}, and other models with \textbf{replica symmetry breaking} \cite{mezard2009information,mezard1987spin}. This is the statistical mechanics terminology for ``lack of self-averaging'' of the order paramaters. In contrast the CW models and high-d inference models in the Bayesian optimal setting are \textbf{replica symmetric}, i.e., the order parameters do concentrate towards their mean as $p\to \infty$ \cite{barbierOverlap,barbierPanchenko}.

\section{Information-theoretic and algorithmic phase transitions}

Until now our discussion was mostly conceptual. But can we practically compute the main high-d quantities we introduced (mutual inference, free entropy, MMSE) in order to understand and predict the behavior of algorithms for high-d inference problems? We continue to focus on the SW model as a representative example, but the following discussion applies more generically.

\subsection{``Single-letter formulas'' for mean-field models: the magic of the concentration of measure}

Deriving single-letter formulas for high-d quantities is often possible for problems belonging to the class of \textbf{mean-field models}. Such formulas usually come in the form of an optimization problem over a function of a scalar parameter. In mean-field models each spin/variable interacts with extensively many other ones, i.e., with $O(p)$: we speak in this case about a \textbf{dense} model. Another class of mean-field models are \textbf{sparse/dilute models}, where the network of interactions between variables is such that in the limit $p\to\infty$, variables $(x_i)$ interact with a random subset of finitely many $O(1)$ other ones. The SW model is a dense mean-field model, as each variable $x_i$ interacts with all the other ones through the pairwise interactions $(\frac12 (y_{ij}-\sqrt{\lambda/p}\,x_ix_j)^2)_{j\le p}$. For such models there exists an arsenal of powerful methods from statistical mechanics that are able to reduce the evaluation of high-d quantities to low-dimensional optimisation problems, in particular the \textbf{replica method} developed in the context of spin glasses \cite{mezard2009information,mezard1987spin}. Such high-d to low-d reduction is another beautiful manifestation of the concentration of measure. 

 \begin{figure}[t!]
    \centering
    \includegraphics[width=10cm]{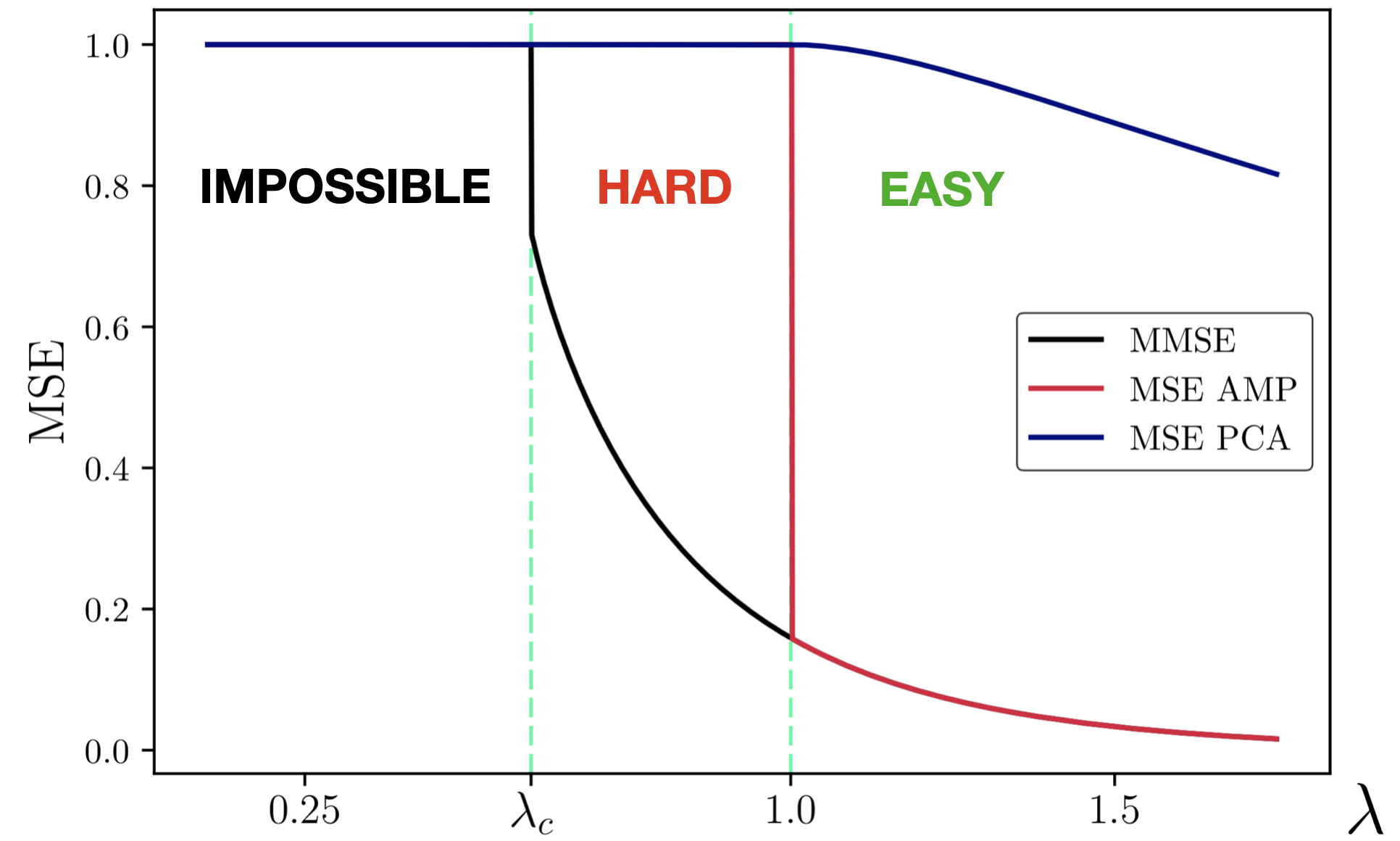}
    \caption{From \cite{leoPHD}. Plot of the spike-MMSE, the MSE of the AMP algorithm and of naive PCA for the SW model with prior \eqref{SBM} with $\rho=0.05$. Information-theoretic and algorithmic first order phase transitions are observed. A computational-to-statistical gap (hard phase) is present between the information-theoretic and algorithmic thresholds.}
    \label{fig:AMPvsPCA}
\end{figure}

Assume again that the prior factorizes with i.i.d. $x_i\sim P$. The replica method (or its close cousin the \textbf{cavity method} \cite{mezard2009information,mezard1987spin}) predicts that the mutual information for the SW model verifies as $p\to\infty$ (denote $v:=\mathbb{E}_P[x^2]$ and $x^*,x$ are i.i.d. from $P$, $z\sim \mathcal{N}(0,1)$ is a standard normal r.v.)
\begin{align*}
\frac1pI(\bx;\by)\!\to\!\! \underset{q\in[0,v]}{\min}\Big\{\frac{\lambda}{4}(q-v)^2+I(x;\sqrt{\lambda q}\, x+z)\Big\}.
\end{align*}
Here $I(x;\sqrt{\lambda q}\, x+z)$ is the mutual information of the Gaussian denoising model with SNR $\lambda q$, given by \eqref{MI_gauss} changing $\lambda$ to $\lambda q$. 
Therefore we can get an actual formula for the mutual information. Equating to $0$ the $q$-derivative of the function $i^{({\rm RS})}(q,\lambda):=\frac{\lambda}{4}(q-v)^2+I(x;\sqrt{\lambda q}\, x+z)$ above --called \textbf{replica-symmetric potential}--, its minimizer $q_{\text{min}}$ verifies the fixed point equation $$\frac \lambda 2(q_{\text{min}}-v)+ \frac{d}{d\lambda}I(x;\sqrt{\lambda q}\, x+z)|_{q=q_{\text{min}}}=0.$$ By the I-MMSE formula it gives
\begin{align}
q_{\text{min}}=v-\text{mmse}(x\mid \sqrt{\lambda q_{\text{min}}}\, x+z)	\label{fp}
\end{align}
where $\text{mmse}(x\mid \sqrt{\lambda q_{\text{min}}}\, x+z)$ is the MMSE for the scalar denoising model; it is given by \eqref{MMSE_denoi} with $\lambda$ replaced by $\lambda q_{\text{min}}$. Whenever unique, the minimizer of the replica symmetric potential can be shown to be equal to $m^*:=\lim_p \mathbb{E}\langle m_p^*\rangle$ (recall \eqref{planted_mag}). Therefore from \eqref{MMSE_ptoMMSE} we also get a ``single-letter formula'' for the MMSE:
\begin{align}
\text{MMSE}=v^2-q_{\text{min}}^2.\label{MMSE_MF}
\end{align}
It is absolutely amazing that such high-d objects, that depend on so many random variables, can be reduced to such simple formulas! There is something very peculiar happening here: both at the level of the mutual information and of the MMSE the simple scalar denoising model appears. The analysis of the high-d SW model therefore collapses onto the analysis of an inference problem of a single signal component corrupted by Gaussian noise, with a SNR $\lambda q_{\text{min}}$ given by a non-trivial fixed point equation. This obervation is generic for dense mean-fields models. For sparse problems things are a bit more subtle but essentially the same type of reduction from a high-d to low-d problems happens too.

Note that we said at some point that the scalar denoising model was not so interesting in itself as there was no phase transition in its MMSE. But here even if this simple model appears, the complexity of the SW is revealed in the fact that the solutions of the fixed point equation \eqref{fp} may be more than one. So from one SNR value $\lambda$ to a close one $\lambda+\varepsilon$, the solution $q_{\text{min}}$ that minimizes the replica symmetric potential (and then gives the MMSE through \eqref{MMSE_MF}) may change discontinuously: a phase transition then occurs.

All these results can be even turned in mathematically rigorous statements. Complementary to the replica method, there exist the so-called \textbf{cavity and interpolation methods} \cite{panchenko2013sherrington,talag1,talag2,guerra2002thermodynamic,guerra2003broken}, applied to the SW model in \cite{lelarge2019fundamental}. Recently an evolution of the interpolation method for high-d inference, called \textbf{adaptive interpolation method}, had great success in proving such fomulas (including the ones given above for the SW model) \cite{barbier2019adaptive,barbier2019adaptive_2,barbier2019optimal}\footnote{There exists also an ``algorithmic approach'' to proving high-d replica symmetric formulas \cite{dia2016mutual,barbier-mutual-2020}.}. For those interested in knowing more about these proof techniques see \cite{barbPisa,leoPHD}.

\subsection{Phase transitions and phase diagram}

\begin{figure}[t!]
    \centering
    \includegraphics[width=10cm]{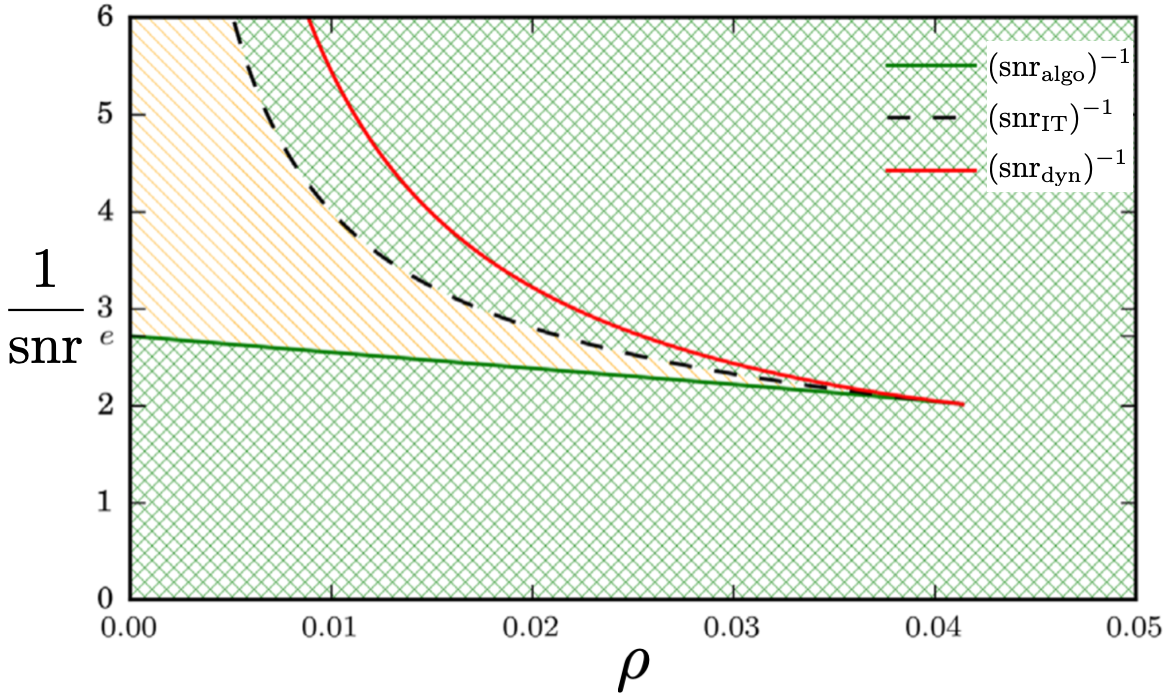}
    \caption{From \cite{lesieur2017constrained}. Phase diagram of the SW model with Bernoulli parameters $x_i\sim \text{ Ber}(\rho)$ as a function of the sparsity $\rho$ and inverse of the total SNR given by $\text{snr}:= \lambda \rho^2$. There is no phase transition in the system if $\rho> 0.0414$ and a first order phase transition else. The lower green curve is the algorithmic phase transition of the AMP algorithm. The dashed black line is the information theoretic threshold. The orange hashed zone is the hard region in which AMP is sub-optimal (as any known sub-exponential complexity algorithm). In the rest of the phase diagram (green hashed) the AMP provides in the large size limit the optimal MMSE.}
    \label{fig:phase_diag_lesieur}
\end{figure}

Equipped with the explicit formula \eqref{MMSE_MF} for the MMSE we are ready to explore the phase diagram of the problem. In Figure~\ref{fig:AMPvsPCA} the MMSE as well as the MSE reached by two algorithms for the SW model is plotted. These algorithms are \textbf{principal component analysis (PCA)} and the \textbf{approximate message-passing (AMP)} algorithm. In PCA one computes the eigenvector of the data matrix $\by$ associated with the maximum eigenvalue; this is the estimator of the signal. Above some \textbf{algorithmic threshold} this eigenvector estimator starts to align with the signal so that the MSE lowers down. We will not discuss the AMP algorithm, but essentially what matters is that it is conjectured by many to be optimal among all low-complexity/practical algorithms in a broad class of high-d inference problems. Here we indeed observe that AMP requires a lower SNR than PCA to perform well (i.e., can perform better at higher noise levels). And when it works it matches the MMSE estimator performance. Moreover its performance in the limit $p\to\infty$ can be rigorously predicted. This allows to get the curves presented here, see \cite{zdeborova2016statistical,lesieur2017constrained,FloCS,barbier2019optimal} for details.

What we observe is a generic scenario in high-d inference with two types of phase transitions delimiting three phases: $i)$ the \textbf{impossible phase} is the regime where even the optimal MMSE estimator performs poorly (not better than random guessing). Therefore it is information-theoretically impossible to infer anything about the signal better than random guessing. It is not a computational issue, there is simply not enough information. The information-theoretic threshold is denoted $\lambda_c$. The SNR regime $\lambda\in (\lambda_c, \lambda_{\text{algo}})$ (where in this problem the algorithmic threshold $\lambda_{\text{algo}}=1$ is the same for PCA and AMP) is the \textbf{hard phase}. Hard is in the algorithmic sense. It means that we do not know any computationally efficient algorithm able to match the performance of the optimal MMSE estimator. Finally $\lambda>\lambda_{\text{algo}}$ corresponds to the \textbf{easy phase}: in this regime we do know a computationally efficient algorithm (AMP) able to match the MMSE. In this model with this specific prior both the information-theoretic and algorithmic transition of AMP are sharp/discontinuous: they are of the first order type. Sometimes they are continuous like hre for the PCA estimate. The precense of an hard phase defines a so-called \textbf{computational-to-statistical gap} (another name for the hard regime), and understanding whether such gap is fundamental or not is one of the main open question in the field. By fundamental we mean whether there actually exists or not in this region a polynomial-time (in $p$) algorithm able to beat AMP and match the MMSE.

\begin{figure}[t!]
\centering
\includegraphics[trim={0 0 0 1.2cm},clip,width=10cm]{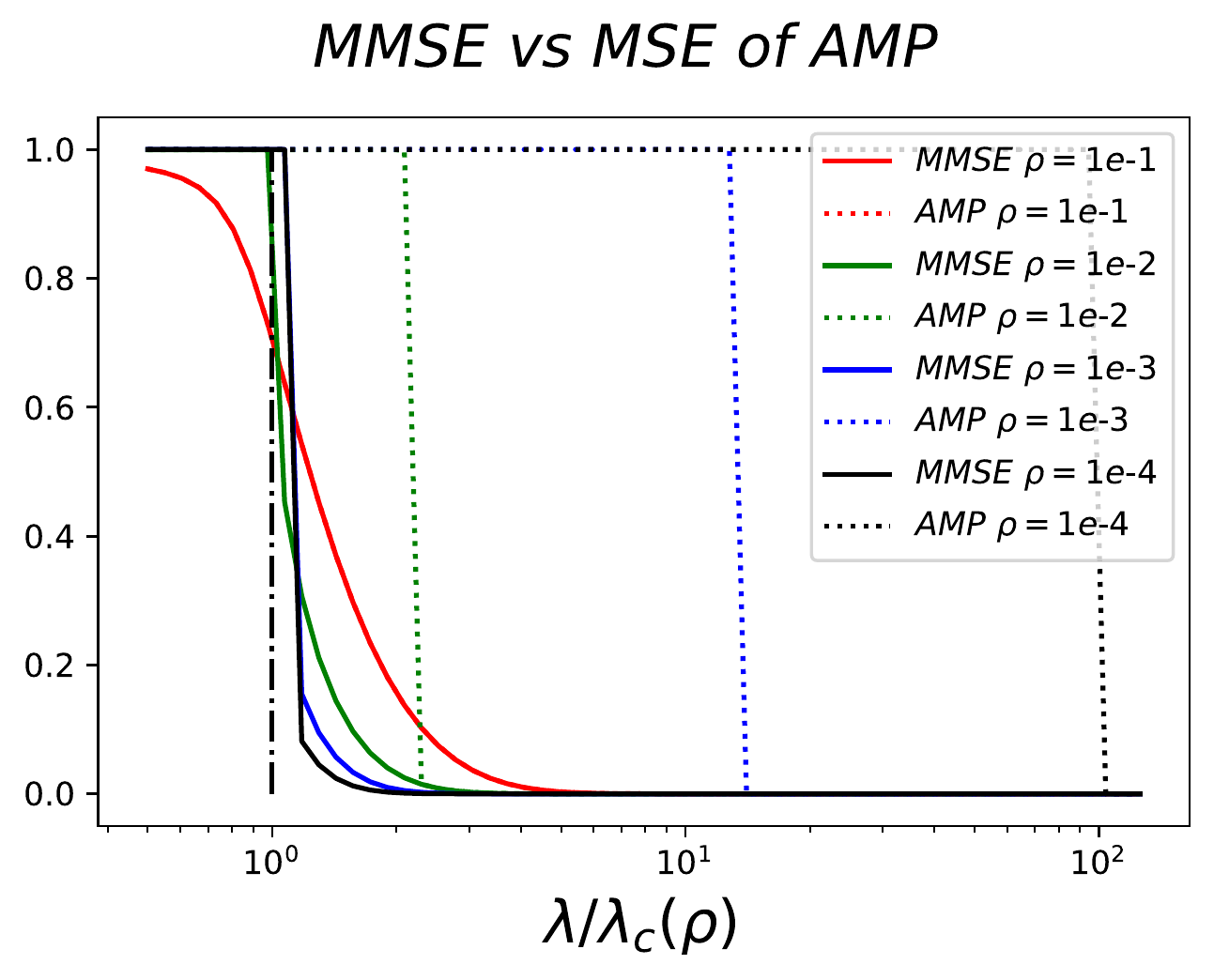}
 \caption{\footnotesize From \cite{barbier2020all}. All-or-nothing information-theoretic and AMP algorithmic phase transitions for the SW model with Bernoulli parameters $x_i\sim \text{Ber}(\rho)$. As the sparsity $\rho$ dereases both transitions become sharper: an all-or-nothing transition appears in the limit $\rho\to0$. Horizontal axis is on a log scale, and is relative to the information-theoretic threshold $\lambda_c(\rho)$, itself function of $\rho$ (see \cite{barbier2020all} for its expression). The statistical-to-algorithmic gap diverges as $\rho\to 0$: it becomes algorithmically harder to infer the signal.}
\label{fig:MMSEandMSEamp}
\end{figure}

These three regimes are separated by phase transitions. Consider the SW model with Bernoulli prior of mean $\rho$. We show the phase transitions lines in the $(1/(\lambda\rho^2),\rho)$ plane (these are the control parameters; $\text{snr}=\lambda\rho^2$ is the natural SNR parameter) in Figure~\ref{fig:phase_diag_lesieur}. Predicting the performance of the MMSE and AMP estimators at each point, it allows to draw the phase diagram of the problem. We observe large regions in green where AMP is optimal, and the hard phase in orange. This is similar to the phase diagram of water in the (temperature, pressure) plane with the solid, liquid and gas phases. This kind of pictures allow to read fundamental and algorithmic limitations of signal reconstruction as control parameters are varied.

Let us mention another interesting observation. It was proven recently in \cite{barbier2020all} (based on conjectures in \cite{lesieur2017constrained}) that \textbf{all-or-nothing phase transitions} happen in the regime of very high sparsity $\rho\to 0$ (still considering a Bernoulli prior for the signal entries). This means that, as observed in Figure~\ref{fig:MMSEandMSEamp}, the transitions become as sharp as they can be in this particular limit. It means that when the \textbf{effective dimension of the signal} is much smaller than its ambient dimension $p$, the signal can be or perfectly infered, or not at all. There is no crossover between these two behaviors like in Figure~\ref{fig:AMPvsPCA} which is for a finite sparsity $\rho$. Here the effective dimension of the signal is $\rho p$, i.e., the expected number of non-zero components. It vanishes when compared to the ambient dimension $p$ as $\rho \to 0$. This phenomenology seems very generic and happens in a broad class of other high-d inference models \cite{luneau2020information}. The success of modern signal processing and machine learning in high-d regimes is believed to be partly due to the structure of the data itself and the fact that even if high-dimensional, it has lower effective dimensionality, that is then exploited by algorithms. Therefore designing and analysing simple models that are tractable and serve as idealized paradigms for this setting is of fundamental interest.

\section{Concluding remarks}

We discussed the modern regime of high-d statistics. Focusing on the spike Wigner model as paradigm of high-d inference, we have shown that inference can be recast in the statistical mechanics language. As in more physical models like spins systems (and virtually any sufficiently complex system) the SW model has phase transitions separating different algorithmic regimes of inference.

For the sake of pedagogy we focused on the SW model. But a large part of the concepts we introduced, the phenomenology we presented and the conlusions we drew are much more general and apply to an extremely large class of inference and learning problems. In order to get a broader view and know about many more examples of high-d inference models that can be treated using the statistical mechanics approach I recommend the excellent review \cite{zdeborova2016statistical}. See also the article \cite{barbier2019optimal}. For mathematically oriented readers see \cite{barbPisa} and \cite{leoPHD}. Classical references are the books \cite{engel2001statistical,nishimori2001statistical}.

We end this article by emphasizing again that the selection of topics and provided references are highly subjective. The field is huge and fastly expanding, and it is a doomed task to cover it all in a finite number of pages. Nevertheless I hope that this contribution may motivate the reader to dive deeper in this fascinating field. A complementary set of references can be found here \cite{refsMont}.



\end{document}